\documentclass[review]{elsarticle}
\usepackage{lineno,hyperref}
\usepackage[english]{babel}
\usepackage[numbers]{natbib}
\usepackage{xcolor}
\usepackage{latexsym,amsmath,amssymb,amsbsy,graphicx,geometry}
\usepackage{epstopdf}
\usepackage{epsfig}
\modulolinenumbers[5]

\begin{document}

\begin{frontmatter}

\title{Direct evaluation of attachment and detachment rate factors of atoms in crystallizing supercooled liquids}

\author[adr1]{Dinar T. Yarullin} 
\ead{YarullinDT@gmail.com}
\author[adr1,adr2]{Bulat N. Galimzyanov} 
\ead{bulatgnmail@gmail.com}
\author[adr1,adr2]{Anatolii V. Mokshin} 
\ead{anatolii.mokshin@mail.ru}

\address[adr1]{Kazan Federal University, 420008 Kazan, Russia}
\address[adr2]{Udmurt Federal Research Center of the Ural Branch of the Russian Academy of Sciences, 426067 Izhevsk, Russia}

\begin{abstract} 
Kinetic rate factors of crystallization have a direct effect on formation and growth of an ordered solid phase in supercooled liquids and glasses. Using crystallizing Lennard-Jones liquid as an example, in the present work we perform a \textit{direct} quantitative estimation of values of the key crystallization kinetic rate factors -- the rate $g^{+}$ of particle attachments to a crystalline nucleus and the rate $g^{-}$ of particle detachments from a nucleus. We propose a numerical approach, according to which a statistical treatment of the results of molecular dynamics simulations was performed without using any model functions and/or fitting parameters. This approach allows one to accurately estimate the critical nucleus size $n_{c}$. We find that for the growing nuclei, whose sizes are larger than the critical size $n_{c}$, the dependence of these kinetic rate factors on the nucleus size $n$ follows a power law. In the case of the subnucleation regime, when the nuclei are smaller than $n_{c}$, the $n$-dependence of the quantity $g^{+}$ is strongly determined by the inherent microscopic properties of a system and this dependence cannot be described in the framework of any universal law (for example, a power law). It has been established that the dependence of the growth rate of a crystalline nucleus on its size goes into the stationary regime at the sizes $n>3n_{c}$ particles. 
\end{abstract}


\end{frontmatter}

\section{Introduction}

Crystallization and condensation are processes in which the rates of attachment and detachment of monomers (atoms, molecules) to and from nuclei play an important role in the nucleation and growth kinetics~\cite{Kelton_Greer_1986,Yasuoka_Matsumoto_1998}. The details of the condensation kinetics have been well studied by experimental and molecular dynamics simulation methods~\cite{Schaaf_Senger_2001,Diemand_Angelil_2013}. Although crystallization of supercooled liquids has also been the subject of extensive studies, the crystallization kinetics is not well understood, especially for deep levels of supercooling~\cite{Sosso_Chen_2016}. One of the main reasons for this is the absence of studies, which are focused on accurate quantification and theoretical description of the monomer gain and loss processes during the crystallization of supercooled liquids and glasses~\cite{Kashchiev_Nucleation_2000}.

The transition rate $g^{+}$ of particles from a liquid to a crystalline phase and the detachment rate $g^{-}$ of particles from a crystalline nucleus are the main kinetic rate factors in the theory of nucleation~\cite{Kelton_Greer_1986, Kashchiev_Nucleation_2000}. These kinetic factors are required to determine the rate characteristics of the crystal nucleation and crystal growth processes~\cite{Agrawal_2014,Mura_Zaccone_2016,Baidakov_2019,Mokshin_Galimzyanov_PCCP_2017,Rodrigues_2018}. Therefore, the quantities $g^{+}$ and $g^{-}$ are included in the master equations of well-known kinetic theories and theoretical models that describe the nucleation and growth of crystals. The Wilson-Frenkel theory~\cite{Wilson_1900,Frenkel_1932}, the Turnbull-Fisher model~\cite{Turnbull_Fisher_1949}, the Kelton-Greer model~\cite{Kelton_Greer_1986} and the gain-loss theory~\cite{Weinberg_2002} are among such the kinetic theories and models.

Direct experimental measurement of the kinetic rate factors $g^{+}$ and $g^{-}$ for a crystallizing bulk system is a very complex task. This is due to difficulties in tracking the trajectories of individual atoms of nano-sized scales. The quantity $g^{+}$ can be calculated indirectly by methods based on Turnbull or Kelton equations~\cite{Turnbull_1961,Kelton_1991,Baidakov_2019,MG_JETPLett_2019} by using experimentally measured self-diffusion coefficient or viscosity. Note that the accuracy of these methods is usually insignificant, especially, for the systems at deep supercooling. On the other hand, classical molecular dynamics simulations are an excellent tool to extract complete information about a crystallizing system as well as for the study of nucleation and growth processes ~\cite{Alder_MD,Orava_Greer_2014,Kirova_Pisarev_2019,Galenko_Salhoumi_2019,Kamaeva_Ryltsev_2020,Ryltsev_Chtchelkatchev_2020,Inogamov_Khokhlov_2020}. In this regard, the results of molecular dynamics simulations can be used to evaluate the kinetic rate factors.

Statistical treatment of information obtained from molecular dynamics simulations can be performed using the mean first-passage time (MFPT) method~\cite{Wedekind_2007, Mokshin_Galimzyanov_2012, Mokshin_Galimzyanov_PCCP_2017}. The MFPT-method is straightforward to implement in a simulation and this method allows one to determine the activation barrier, growth curves, and lag times~\cite{Wedekind_2007}. As shown before, this method can be used to estimate the value of the kinetic rate factor $g_{n_{c}}^{+}$ for the critically-sized nucleus [Ref.~\cite{Wedekind_2008,Lundrigan_2009,Mendelev_2018}]. Another method proposed by Auer and Frenkel~\cite{Auer_Frenkel_2004} to compute the quantity $g_{n_{c}}^{+}$ for nuclei with the critical size $n_{c}$ is also based on the statistical treatment of trajectories of growing crystalline nuclei. In this method, it is assumed that the size of a nucleus fluctuates around its critical value and this nucleus grows via the diffusive attachment of single particles. 

In the present work, we propose a simple and accurate approach for direct evaluation of the kinetic rate factors $g^{+}(n)$ and $g^{-}(n)$. According to this approach, the calculations can be performed as the size-dependent quantities without using model functions and fitting parameters. We demonstrate the efficiency of the approach for the case of study of the crystallization kinetics of supercooled Lennard-Jones (LJ) liquid.

\section{Approach for direct evaluation of the kinetic rate factors}

Let us consider an idealized situation of mononuclear crystallization when a single crystalline nucleus grows isotropically in a supercooled liquid. The growth of this nucleus occurs due to the local rearrangements of parent (disordered) phase particles, which are located near the surface of a crystalline nucleus. Then, the quantity $k^{+}$ will determine the number of particles (monomers) attached to the nucleus surface in a unit time. At the same time, a crystalline nucleus can decay due to the detachment of particles from its surface. The number of detached particles we denote as $k^{-}$. Thus, a nucleus grows when $k^{+}>k^{-}$. On the other hand, the nucleus size remains unchanged when $k^{+}=k^{-}$; whereas the nucleus size decreases in the case $k^{+}<k^{-}$.

\begin{figure}
	\centering
	\includegraphics[width=0.65\linewidth]{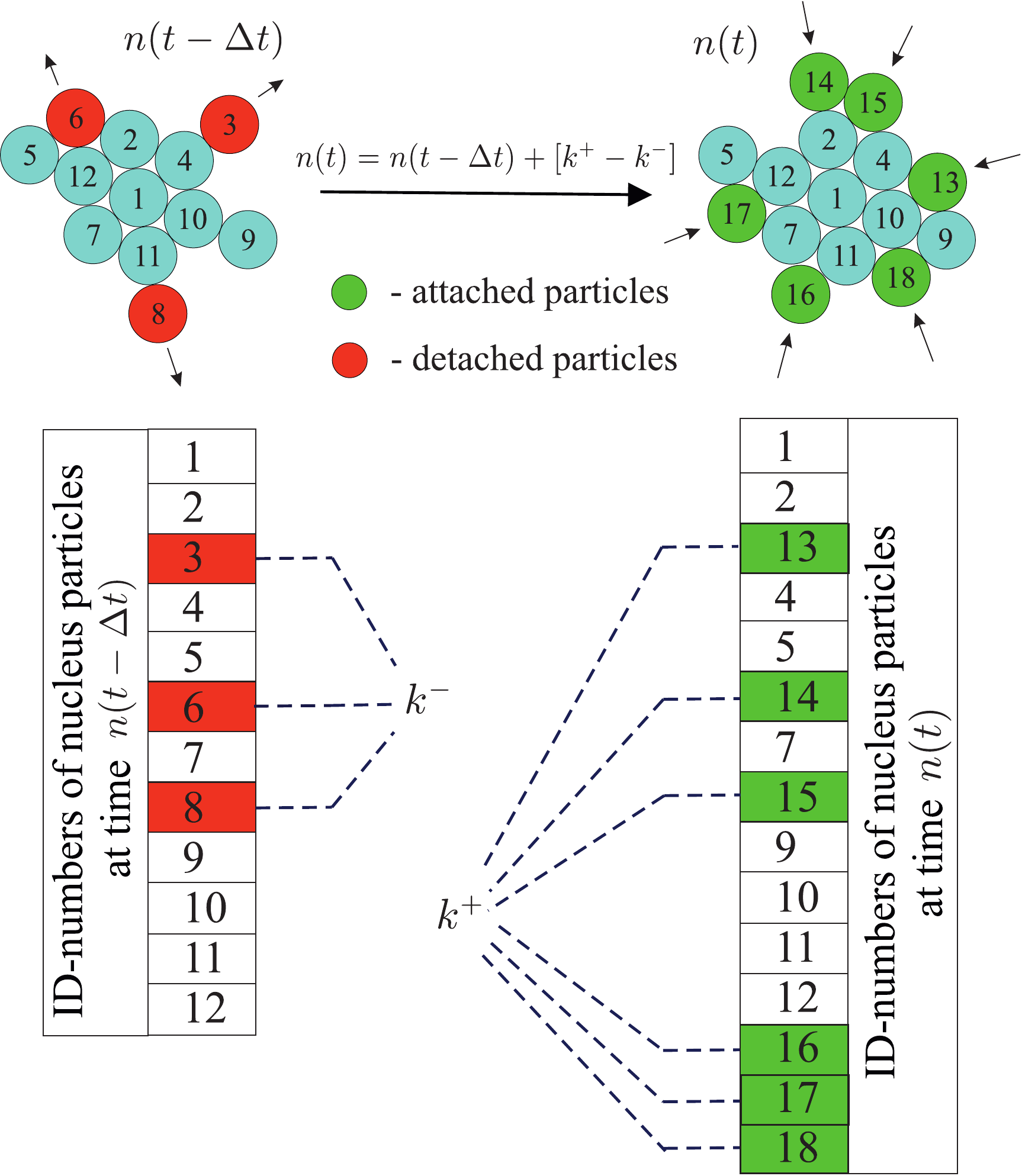}
	\caption{Schematic procedure for evaluation of the quantities $k^{+}$ and $k^{-}$ taking into account the identification (label) numbers of the particles of a crystalline nucleus. Particles detached from the nucleus surface are in red, whereas the attached particles are colored green.}
	\label{fig_1}
\end{figure}

Let us suppose that the trajectories  $\vec{r}_{i}(t)$ [$i=1,2,...,N$, $N$ is the number of particles] of all particles in the system are known. Each particle has a unique label number, which is assigned during cluster analysis of simulation results. Then, a nucleus of the size $n(t)$ at the time $t$ can be represented as a one-dimensional array that consists of the labels of nucleus particles. Changes of label numbers in this array are tracked at each simulation time step [see scheme in Fig.~\ref{fig_1}]. So, the appearance of new labels (particles) in this array is the manifestation of the so-called gain process. The number of such the labels will determine the value of the quantity $k^{+}(n)$. The disappearance of labels from the array corresponds to the loss process and the number of such the labels defines value of the quantity $k^{-}(n)$.

The mononuclear crystallization scenario considered above is very specific and it implements rarely. The polynuclear scenario is more common at the crystallization of supercooled liquids and glasses. This scenario involves formation of stable crystalline nuclei in a system~\cite{JETP_2018}. In the case of a high concentration of crystalline nuclei, these nuclei grow mainly due to coalescence with each other~\cite{JCG_2019}. Therefore, for evaluate the quantity $k^{+}$ we take into account only the particles, which transfer to a the crystalline phase directly from a disordered phase. On the other hand, a crystalline domain can decay into separate parts when incomplete coalescence of nuclei occurs or when a nucleus has a highly ramified shape. Such the decay can lead to a rapid increase in the number of detached particles, $k^{-}$. Therefore, at calculation the value of the quantity $k^{-}$ we consider the particles that transfer from the crystalline phase to the disordered phase. Exclusion from consideration of the nuclei coalescence and nucleus decay processes allows one to determine values of the kinetic rate factors $g^{+}$ and $g^{-}$ with high accuracy. Thus, the proposed approach is not limited to any supercooling regime since these kinetic rate factors are calculated from their basic definitions.

According to the basic definition~\cite{Kashchiev_Nucleation_2000}, the kinetic rate factor $g^{+}$ characterizes the number of particles (monomers) attached to a crystalline nucleus of size $n$ over the shortest time step $\Delta t$. The kinetic rate factor $g^{-}$ determines the number of particles detached from $n$-sized crystalline nucleus and transferred to the parent phase over the time step $\Delta t$. Thus, the expressions for estimation the values of the quantities $g^{+}$ and $g^{-}$ will have the following forms 
\begin{equation}\label{eq_basic_gp}
g^{+}(n)=\frac{\left\langle k^{+}(n)\right\rangle}{\Delta t},
\end{equation}
\begin{equation}\label{eq_basic_gm}
g^{-}(n)=\frac{\left\langle k^{-}(n)\right\rangle}{\Delta t}.
\end{equation}
In the present work, the simulation time step is $\Delta t = 0.01\,\tau$ [in the case of argon with the parameters $m=6.63\times10^{-26}$~kg, $\sigma=0.341$~nm, $\epsilon/k_{B}=119.8$~K and $\tau=\sigma\sqrt{m/\epsilon}$, this time step is $\simeq0.0215$~ps]. The brackets $\langle...\rangle$ denote averaging over various molecular dynamics iterations. Statistical treatment of our results is carried out over $50$ independent trajectories $n(t)$ of the largest growing crystalline nucleus.

\section{Results and Discussion}

\subsection{Estimation of the nucleus critical size}

\begin{figure*}
	\centering
	\includegraphics[width=1.0\linewidth]{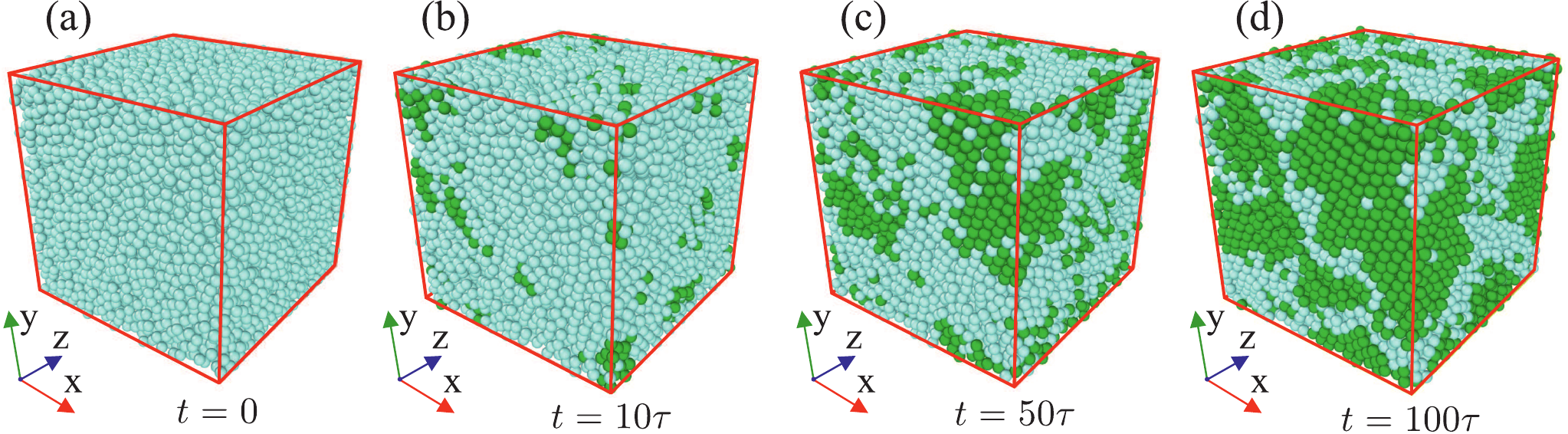}
	\caption{Snapshots of the system: (a) at the time $t=0$, (b) $t=10\,\tau$, (c) $t=50\,\tau$ and (d) $t=100\,\tau$. Light blue circles indicate the particles that form the parent (disordered) phase. The particles involved in the formation of the crystalline phase are indicated by dark green circles.}
	\label{fig_2}
\end{figure*}

In the present work, we consider the crystallization of the supercooled Lennard-Jones liquid at the temperature $T=0.5\,\epsilon/k_{B}$ and the pressure $p=2.0\,\epsilon/\sigma^{3}$. The simulation details and applied methods are given in the Appendix. A well-known peculiarity of one-component LJ-system is that this system is a poor glass-former and it crystallizes rapidly after cooling below the melting temperature $T_{m}$ [Ref.~\cite{Baidakov_Protsenko_2019, Stephan_Thol_2019}]. This means that it is possible to observe the crystal nucleation and crystal growth processes in this crystallizing system at time scales available for simulation. For the considered thermodynamic ($p$, $T$)-state, the stable crystalline nuclei appear at the times $t>10$~$\tau$ [see Figs.~\ref{fig_2}(a) and~\ref{fig_2}(b)]. The high concentration of crystalline nuclei leads to their coalescence at the times $t>50\,\tau$ [see Fig.~\ref{fig_2}(c)]. Fig.~\ref{fig_2}(d) shows that the system forms a polycrystal that is the typical final structure appeared due to the crystallization of a liquid of moderate supercooling levels~\cite{JETP_2018}. To minimize the effects related with the nuclei coalescence and nucleus decay processes on values of the kinetic rate factors, we will consider the nuclei that grow only within the time range $t\in[0;\,50]\,\tau$.

\begin{figure}
	\centering
	\includegraphics[width=0.65\linewidth]{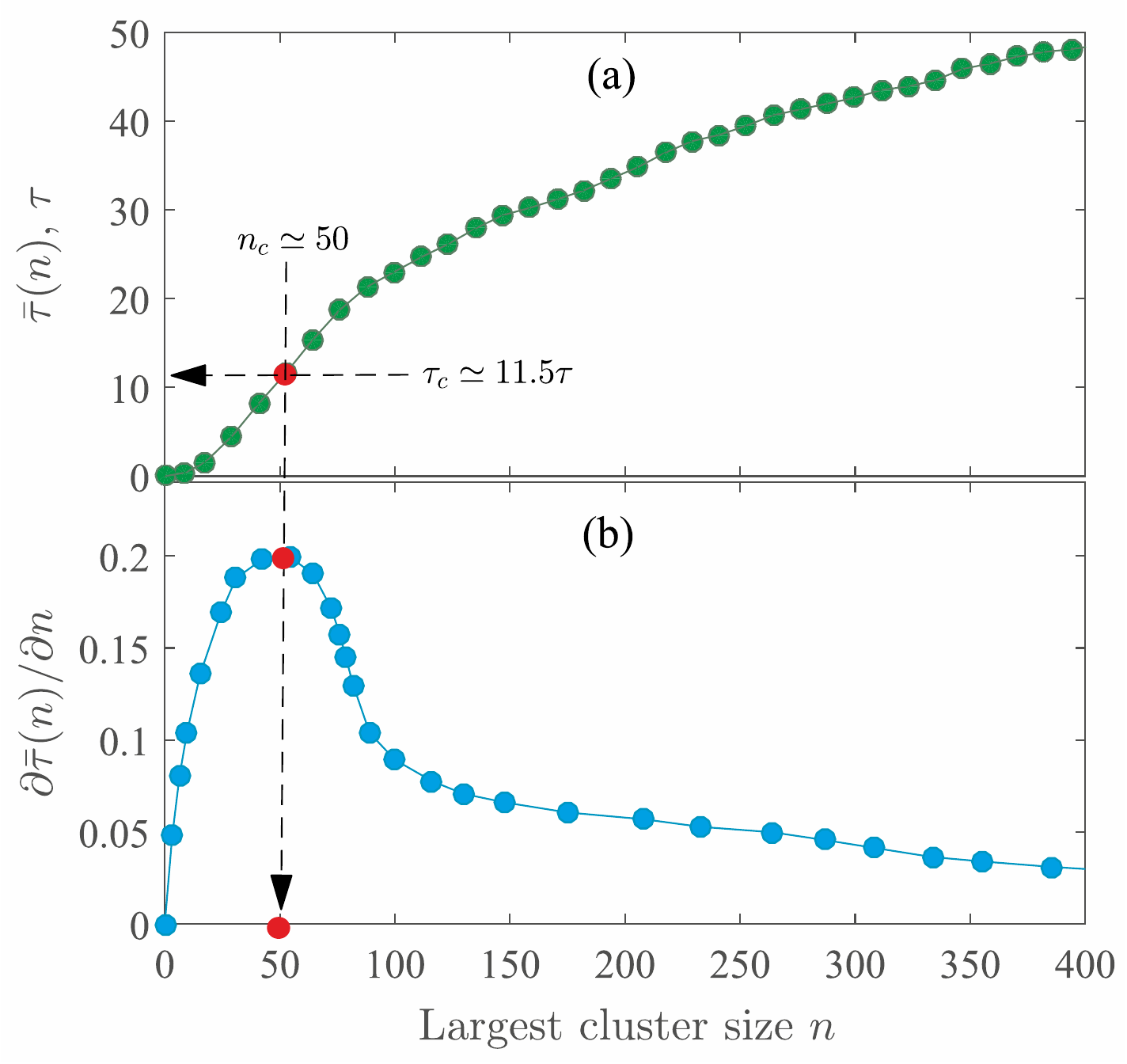} 
	\caption{(a) Mean first-passage time $\bar{\tau}$ of the largest nucleus as a function of the nucleus size $n$. (b) Derivative of the function $\bar{\tau}(n)$. The arrows show the position of the inflection point in the curve $\bar{\tau}(n)$ and the location of the maximum in the function $\partial\bar{\tau}(n)/\partial n$.}
	\label{fig_3}
\end{figure}

Fig.~\ref{fig_3}(a) shows the dependence of the mean first-passage time $\bar{\tau}$ of the largest crystalline nucleus on its size $n$ [Ref.~\cite{Mokshin_Galimzyanov_2012, Mokshin_Galimzyanov_PCCP_2017}]. As can be seen from Fig.~\ref{fig_3}(a), the curve $\bar{\tau}(n)$ contains the pronounced inflection point~\cite{Wedekind_2008, Lundrigan_2009, Gunawardana_2018}. The derivative of $\bar{\tau}(n)$ over variable $n$ has one pronounced maximum, as seen from Fig.~\ref{fig_3}(b). From the location of this maximum on $n$-scale, we find the critical nucleus size $n_{c}\simeq(50\pm3)$ particles. This value of the critical size is typical for spontaneously crystallizing LJ-system at the considered ($p$, $T$)-state~\cite{Baidakov_2019}. The average waiting time for the critically-sized nucleus is $\tau_{c}\simeq(11.5\pm1.2)\,\tau$ [in the case of argon with the interatomic potential parameter $\sigma=0.341$~nm and $\epsilon/k_{B}=119.8$~K, this time $\tau_{c}$ corresponds to the value $(24.7\pm2.58)$~ps]. Such the relatively small value of the waiting time $\tau_{c}$ indicates on an extremely high crystal nucleation rate in the supercooled LJ-system~\cite{Baidakov_Protsenko_2019}. 

Based on the obtained MFPT-curve shown in Fig.~\ref{fig_3}, we have calculated the height of the nucleation barrier~\cite{Wedekind_2007,Mokshin_Galimzyanov_2012}
\begin{equation}\label{eq_mfpt_gibbs_eng}
\beta\Delta G^{*}=\frac{3\pi}{4}\left(\frac{n_{c}}{\tau_{c}}\right)^{2}\left[\frac{\partial\bar{\tau}(n)}{\partial n}\Bigg|_{n=n_{c}}\right]^{2},
\end{equation}
which takes the value $\beta\Delta G^{*}\approx(2.6\pm0.4)$ (where $\beta=(k_{B} T)^{-1}$) for the LJ-system in the considered ($p$, $T$)-state. In our study, the system at the high pressure $p=2.0~\epsilon/\sigma^3$ is considered. As we know, high pressures can accelerate crystallization~\cite{Wolde_Frenkel_1996,Koperwas_Affouard_2017}. As a result, we have such the small nucleation barrier, which provides the fact that nucleation events occur in time scales available for molecular dynamics simulation.

\subsection{Size-dependence of the attachment rate}

\begin{figure*}
	\centering
	\includegraphics[width=1.0\linewidth]{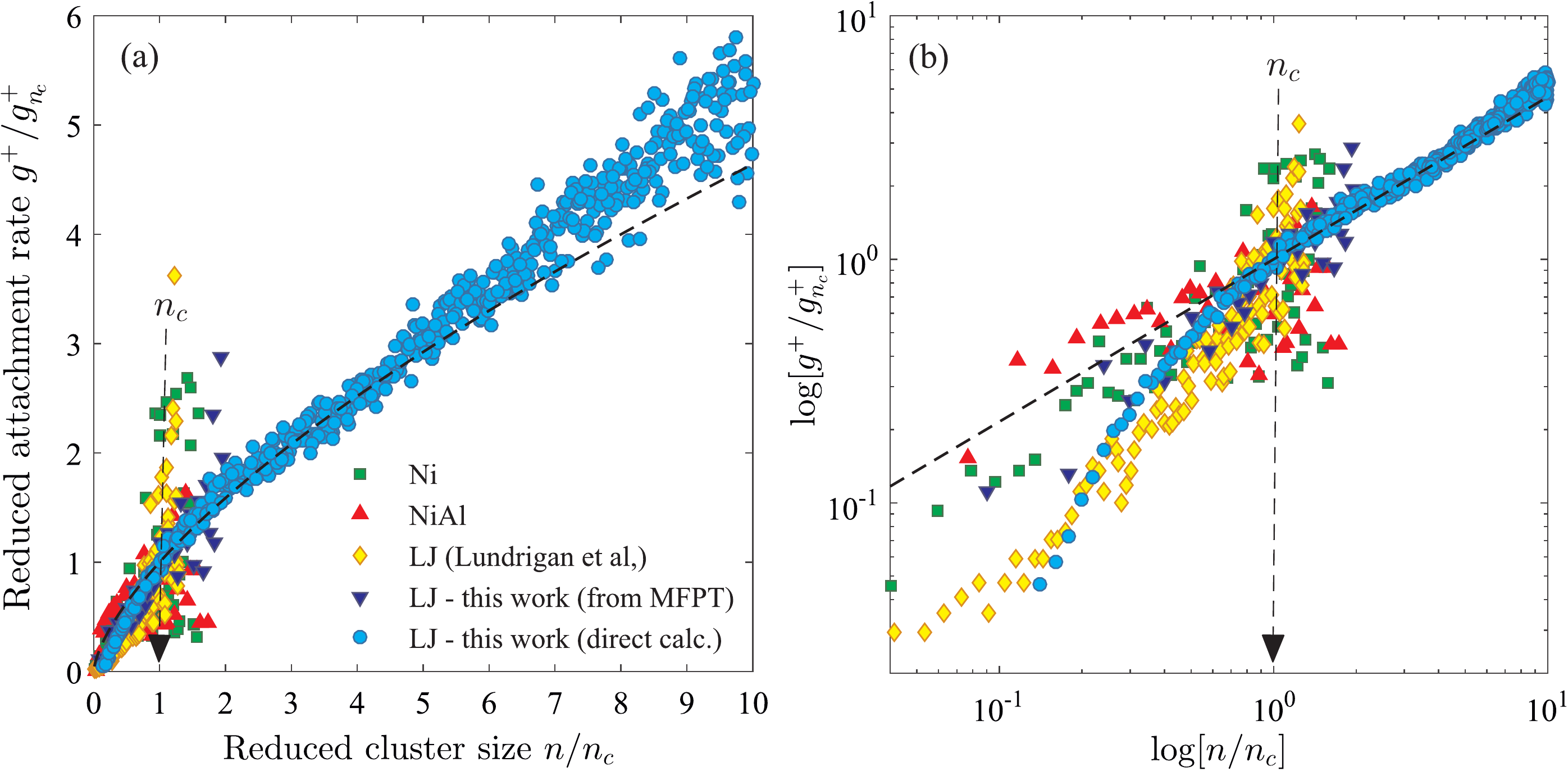}
	\caption{(a) Reduced kinetic rate factor $g^{+}/g^{+}_{n_c}$ as a function of the reduced nucleus size $n/n_c$ for crystallizing LJ-system at the supercooling $\Delta T/T_{m}=0.43$. Our results are compared with simulation data obtained for crystallizing LJ-system at the supercooling $\Delta T/T_{m}=0.49$ [Ref.~\cite{Lundrigan_2009}] as well as for crystallizing pure Ni and binary NiAl alloy at the supercooling $\Delta T/T_{m}=0.3$ [Ref.~\cite{Mendelev_2018}]. The wide scatter in values of the kinetic rate factor $g^{+}$ at the size $n\approx n_{c}$ is due to this quantity is evaluated near the maximum of the nucleation barrier at which the nucleus is unstable. (b) $\log[g^{+}/g^{+}_{n_c}]$ vs. $\log[n/n_c]$ curves for the considered systems. The dashed curve is obtained by Eq.~(\ref{eq_powerlaw}) at $\xi=1$.}
	\label{fig_4}
\end{figure*}

Fig.~\ref{fig_4}(a) shows the dependence of the reduced kinetic rate factor $g^{+}/g^{+}_{n_{c}}$ on the reduced nucleus size $n/n_{c}$, where $g^{+}_{n_{c}}$ is the rate of particles attachment to the critically-sized nucleus $n_{c}$ [see Table~\ref{tab_1}]. In this dependence, two regimes can be distinguished, which are typical for the quantity $g^{+}(n)$  [Ref.~\cite{Kashchiev_Nucleation_2000}]. The first regime covers the size range $n/n_{c}\in[0; 1]$ corresponding to subcritically-sized crystalline nuclei. The second regime at the sizes $n/n_{c}>1$ corresponds to growing supercritical nuclei.
\begin{table}
	\centering
	\caption{Parameters of the considered systems: supercooling $\Delta T/T_{m}$, melting temperature $T_{m}$, critical size $n_{c}$, particles attachment rate $g^{+}_{n_{c}}$ to critically-sized nucleus. The attachment rate $g_{n_{c}}^{+}$ calculated for Ni and NiAl is presented in reduced units using the Lennard-Jones parameters from Refs.~\cite{Heinz_Vaia_2008,Sanati_Albers_2001}.}
	\begin{tabular}{cccccc}
		\hline\hline
		System & $\Delta T/T_{m}$ & $T_{m}$ & $n_{c}$ & \,\, & $g_{n_{c}}^{+}$ \\
		\hline
		LJ (our work)   		  & $0.43$ & $0.88\,\epsilon/k_B$ & $50$  &  \,\, & $488\,\tau^{-1}$  \\
		LJ (from Ref.~\cite{Lundrigan_2009})  & $0.49$ & $1.15\,\epsilon/k_B$ & $71$  &  \,\, & $43\,\tau^{-1}$  \\
		Ni (from Ref.~\cite{Mendelev_2018})   & $0.3$ & $1728$~K & $52$  &  \,\, & ($1.07\times10^{14}$)~s$^{-1}$ [$\simeq21.4\,\tau^{-1}$]  \\
		NiAl (from Ref.~\cite{Mendelev_2018}) & $0.3$ & $1821$~K & $26$  &  \,\, & ($1.26\times10^{12}$)~s$^{-1}$ [$\simeq0.32\,\tau^{-1}$] \\
		\hline\hline
	\end{tabular}\label{tab_1}
\end{table}
The found ($n/n_{c}$)-dependence of the reduced kinetic rate factor $g^{+}/g^{+}_{n_ {c}}$ is compared with the simulation data obtained for LJ-system at the supercooling $\Delta T/T_{m}=0.49$ [at the number density $\rho=0.95\,\sigma^{-3}$]~\cite{Lundrigan_2009}, as well as with the data obtained for pure Ni and binary NiAl-alloy at the supercooling $\Delta T/T_{m}=0.3$ [Ref.~\cite{Mendelev_2018}]. The values of the critical size $n_{c}$ and the kinetic rate factor $g_{n_{c}}^{+}$ estimated for the considered systems are shown in Table~\ref{tab_1}. As seen from Fig.~\ref{fig_4}(a), all the dependencies are similar and have a common trend: the larger the nucleus size $n$, the greater the attachment rate $g^{+}$ of particles to the nucleus surface. Here, the values of the kinetic rate factor $g^{+}$ estimated by our approach are obtained for nuclei with sizes up to $10n_{c}$ particles. As far as we know, the quantity $g^{+}$ has not been previously evaluated for nuclei with sizes more than $2n_{c}$. 

In addition, we have computed the attachment rate by the mean first-passage time $\tau(n)$ using Wedekind and Reguera method~\cite{Wedekind_2008}:
\begin{equation}\label{eq_mfpt_gplus_1}
g^{+}(n)=B(n)/\frac{\partial\bar{\tau}(n)}{\partial n},
\end{equation}
where
\begin{equation}\label{eq_mfpt_gplus_2}
B(n)=-\frac{1}{P(n)}\left[\int_{n}^{2n_{c}}P(n')dn'-\frac{2\tau_{c}-\bar{\tau}(n)}{2\tau_{c}}\right].
\end{equation}
Here $P(n)$ is the probability of the formation of the largest nucleus with size $n$ in the system. As seen from Fig.~\ref{fig_4}, the approach for direct estimation and Eq.~(\ref{eq_mfpt_gplus_1}) have a similar tendency to increase $g^{+}$ with increasing nucleus size. The agreement between the two approaches is good despite some noise from Eq.~(\ref{eq_mfpt_gplus_1}). The $n$-dependence of the quantity $g^{+}$ found by Eq.~(\ref{eq_mfpt_gplus_1}) is also in good agreement with the data of Lundrigam’s et al. obtained for crystallizing LJ-system with the supercooling $\Delta T/T_{m}=0.49$ [Ref.~\cite{Lundrigan_2009}].

The ($n/n_{c}$)-dependencies of the kinetic rate factor $g^{+}$ shown on Fig.~\ref{fig_4}(a) can be well fitted by the power-law~\cite{Mokshin_Galimzyanov_PCCP_2017}
\begin{equation}\label{eq_powerlaw}
g^{+}(n)=g_{n_{c}}^{+}\left(\frac{n}{n_{c}}\right)^{(3-\xi)/3},\,\,0<\xi\leq3.
\end{equation} 
Here, the exponent $\xi$ characterizes the growth regime of crystalline nuclei. If the exponent $\xi=3$, then the quantity $g^{+}$ is independent of the nucleus size, whereas at the value $\xi=0$ this kinetic rate factor changes with increasing nucleus size according to $g^{+}(n)\sim n$. For LJ-system at the considered ($p$, $T$)-state, the value of the exponent is $\xi\simeq1.0$. This value indicates that the attachment rate is proportional to the nucleus surface~\cite{Volterra_Cooper_1985}. This corresponds to the so-called ballistic model with $g^{+}(n)\sim n^{2/3}$ in Eq.~(\ref{eq_powerlaw})~\cite{Mokshin_Galimzyanov_PCCP_2017}. In particular, this scenario is most often realized in the case of fluid droplet growth in a supersaturated vapor~\cite{Volterra_Cooper_1985,Diemand_Angelil_2013}. As we found, the power-law (\ref{eq_powerlaw}) is valid only for nuclei with sizes $n_{c}<n\leq5n_{c}$ that is seen from Fig.~\ref{fig_4}(a). From the results of cluster analysis, it follows that the nuclei with the sizes $n>5n_{c}$ interact with other small nuclei. The largest nucleus often grows through the mechanism of restructuration and absorption of a small nucleus. In particular, such a mechanism was observed during the crystallization of single-component metal melts at low and deep levels of supercooling~\cite{JCG_2019}. Due to the small difference in the structure and free energy of the absorbing and absorbed nuclei, the attachment of particles of the absorbed nucleus occurs faster than the transition of particles from the parent phase. Note that this coalescence mechanism is not excluded from consideration in the proposed approach for the evaluation of kinetic rate factors. Thus, the deviation of the size dependence of the attachment rate from the power-law $\sim n^{2/3}$ for the nuclei with the sizes $n>5n_{c}$ is mainly due to that the largest nucleus interacts with small nuclei. Fig.~\ref{fig_4}(b) shows that at the sizes $n<n_{c}$, the dependence between $\log[g^{+}/g_{n_{c}}^{+}]$ and $\log[n/n_{c}]$ deviates from the straight line with the slope $\xi\simeq1.0$. The $n$-dependence of the quantity $g^{+}$ calculated for subcritical sized nuclei strongly depends on the type of the system, and this dependence is not reproducible by a power law of form (\ref{eq_powerlaw}).

\subsection{Competition between particle attachment and detachment processes}

\begin{figure*}
	\centering
	\includegraphics[width=1.0\linewidth]{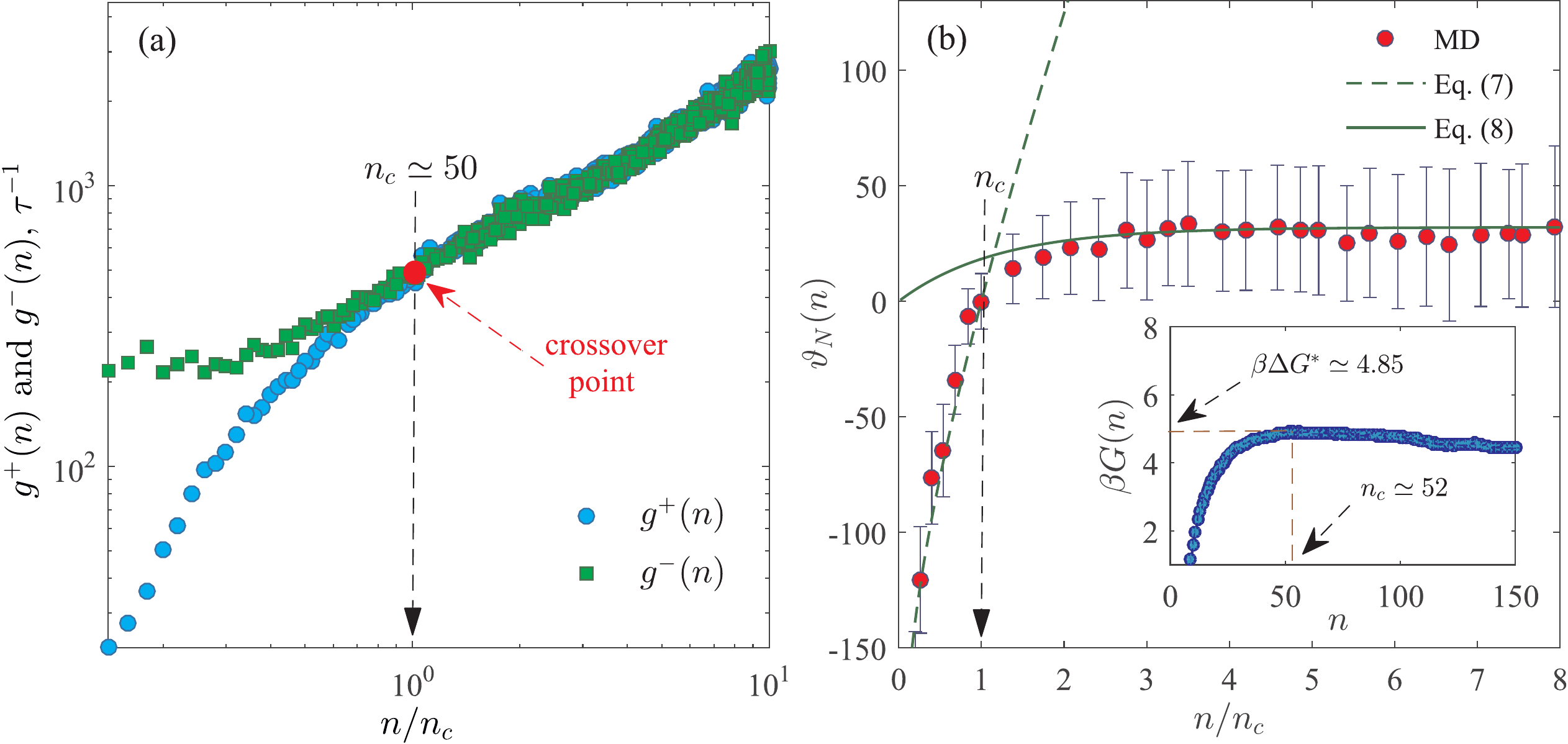}
	\caption{(a) Particles attachment rate $g^{+}$ and particles detachment rate $g^{-}$ as functions of the reduced nucleus size $n/n_{c}$. (b) Dependence of the growth rate $\vartheta$ on the reduced size $n/n_ {c}$ calculated by Eq.~(\ref{eq_growthrate}). The dashed and solid curves are the theoretical results obtained from Eq.~(\ref{eq_model_1}) and Eq.~(\ref{eq_model_2}), respectively. Inset: size-dependence of the free energy $\beta\Delta G(n)$ obtained by Eq.~(\ref{eq_gfe_1}).}
	\label{fig_5}
\end{figure*}

Fig.~\ref{fig_5}(a) shows the ($n/n_{c}$)-dependences of the kinetic rate factors $g^{+}$ and $g^{-}$ in log-log scale.
As can be seen, the crossover of these dependences occurs in the neighborhood of the critical size $n_{c}\simeq50$ particles. Presence of such the crossover point is in agreement with the general ideas of the classical nucleation theory~\cite{Kashchiev_Nucleation_2000}. Obviously, for growing crystalline nuclei, the value of the kinetic rate factor $g^{+}$ will prevail over the value of the quantity $g^{-}$. Then, the difference between $g^{+}$ and $g^{-}$ will determine the growth rate of nuclei in terms of the number of particles~\cite{Mydlarz_Jones_1993}
\begin{equation}\label{eq_growthrate}
\vartheta(n)=g^{+}(n)-g^{-}(n).
\end{equation}
The $n$-dependent growth rate $\vartheta$ is negative for subcritically-sized nuclei, whereas the growth rate $\vartheta$ takes positive values for nuclei with the critical and supercritical sizes [see Fig.~\ref{fig_5}(b)]. The found $n$-dependence of the quantity $\vartheta$ reaches saturation at the sizes $n>3n_{c}$. The growth of nuclei with the sizes $n>3n_{c}$ occurs due to predominance of the kinetic rate factor $g^{+}$ over the quantity $g^{-}$. This finding is in agreement with the data obtained before for crystallizing Dzugutov system and binary LJ-system, where the transition to stationary growth regime occurs at the nucleus size $n\simeq[1.7; 3.0]n_{c}$ [Ref.~\cite{Mokshin_Galimzyanov_PCCP_2017}]. As a rule, such the stable growth ceases either after transition of the crystallization process to the coalescence regime or when the system is fully crystallized~\cite{JETP_2018,JCG_2019}.

The found ($n/n_c$)-dependence of the growth rate $\vartheta$ is reproduced by equation~\cite{Mokshin_Galimzyanov_PCCP_2017}
\begin{equation}\label{eq_model_1}
\vartheta(n)=g^{+}_{n_c}\left(\frac{n}{n_c}\right)^{2/3}\left\{ 1-\exp\left(-\beta\Delta\mu\left[\left(\frac{n}{n_{c}}\right)^{1/3}-1\right]\right)\right\}.
\end{equation}
In Eq.~(\ref{eq_model_1}), the quantity $\Delta\mu$ is the difference between the chemical potentials of the disordered and crystalline phases. As seen from Fig.~\ref{fig_5}(b), the theoretical curve obtained from Eq.~(\ref{eq_model_1}) reproduces only the subcritical nuclei appearance regime, when we use $\beta\Delta\mu\approx(0.85\pm0.1)$. The estimated value of the difference between the chemical potentials is $\Delta\mu\approx0.43~\epsilon$ for LJ-system at the considered ($p$,$T$)-state. For comparison, this value is close to the value $\Delta\mu\approx0.36~\epsilon$ calculated before for the supercooled LJ-system under close thermodynamic conditions~\cite{Gunawardana_2018,Pedersen_2013}. In the saturation regime, the ($n/n_c$)-dependence of the growth rate $\vartheta$ is reproduced by the exponential function
\begin{equation}\label{eq_model_2}
\vartheta(n)=\vartheta_{st}\left\{ 1-\exp\left(-a\left[\frac{n}{n_{c}}\right]\right)\right\}.
\end{equation}
The growth model (\ref{eq_model_2}) was proposed earlier by Mydlarz and Jones to describe the crystal growth in a stable growth regime (see Ref.~\cite{Mydlarz_Jones_1993}). As follows from this growth model, the growth rate is close to the constant value $\vartheta_{st}\approx32~\tau^{-1}$ for nuclei with the sizes $n>n_c$. In the case of argon, this value of $\vartheta_{st}$ is $14.8$~ps$^{-1}$. It is interesting to note that the value of the empirical parameter $a$ coincides with the value of the chemical potential, $a=\beta\Delta\mu\approx0.85$. This indicates that for nuclei with the sizes $n>n_c$, the nature of the relationship between the growth rate $\vartheta$ and the nucleus size $n$ is completely determined by the difference of the chemical potentials $\Delta\mu$.

Inset on Fig.~\ref{fig_5}(b) shows the size-dependence of the Gibbs free energy $\beta\Delta G(n)$ obtained at known values of the kinetic rate factors $g^+$ and $g^-$. We use the expression~\cite{Kelton_Greer_2010}
\begin{equation}\label{eq_gfe_1}
\beta\Delta G(n)=-\int_{1}^{n}\ln\left[\frac{g^{+}(n')}{g^{-}(n'+1)}\right]dn',
\end{equation}
which is obtained from the detailed balance condition
\begin{equation}\label{eq_gfe_2}
g^{-}(n+1)N^{eq}(n+1)=g^{+}(n)N^{eq}(n)
\end{equation}
and Gibbs distribution for crystalline nucleus sizes
\begin{equation}\label{eq_gfe_3}
N^{eq}(n)=N_{0}\exp[-\beta\Delta G(n)].
\end{equation}
Here  $\Delta G(n)$ is the work required to form the $n$-sized nucleus; $N_{0}$ is the pre-exponential constant; $N^{eq}$ is the equilibrium size distribution for the largest nucleus. The position of the maximum in the $n$-dependence of the quantity $\beta\Delta G$ corresponds to the critically-sized nucleus containing $52$ particles. This value almost coincides with the critical size calculated through the derivative of the MFPT-curve (see Table~\ref{tab_1}). Moreover, Eq.~(\ref{eq_gfe_1}) yields the nucleation barrier $\beta\Delta G^{*}\approx(4.85\pm0.35)$  that is comparable with the value $\beta\Delta G^{*}\approx(2.6\pm0.4)$ found also through the MFPT analysis.

\section{Concluding Remarks}\label{s_concl}

Thus, direct evaluation of the kinetic rate factors $g^{+}$ and $g^{-}$ was performed for crystalline nuclei that grow in supercooled Lennard-Jones liquid. The size-dependences of these kinetic rate factors were determined. Calculations were performed for nuclei with sizes up to $10n_{c}$ without using model functions and fitting parameters. As far as we know, such the calculations have not been performed before. The scaled kinetic rate factors were applied to compare our results with other known simulation data obtained for supercooled LJ-liquid as well as for supercooled Ni and NiAl melts. We found that the ($n/n_{c}$)-dependence of the reduced kinetic rate factor $g^{+}/g^{+}_{n_c}$ follows a power law at the nucleus sizes $n\geq n_{c}$. It has been found that in the case of subcritically-sized nuclei (for $n<n_{c}$) this power law is not observed. This finding indicates on a mixing different nucleus growth regimes. Moreover, the dependence of the nucleus growth rate $\vartheta$ on the nucleus size $n$ was calculated. In this dependence, the transition to the stationary growth regime occurs for the nuclei with the sizes $n>3n_{c}$ particles. These results are in good agreement with theoretical calculations.

In conclusion, we note that the results of the present work can be applied to solve the following topical tasks: development of more accurate methods for evaluation of the rate characteristics of structural transformations in systems with different physical and chemical properties (ionic liquids, molecular liquids, polymer systems, colloidal solutions)~\cite{Huang_Ruan_2017,Khusnutdinoff_2019,Brazhkin_2019}; development of severe model and/or theory for describe the nucleus size dependence of the kinetic rate factors $g^{+}$ and $g^{-}$; quantitative characterization and theoretical description of the decay process of crystalline structures in supercooled liquids and glassy materials.

\section*{Acknowledgement}
\noindent This work is supported by the Russian Science Foundation (project 19-12-00022).

\section*{Appendix: {Parameters of the considered system and applied methods}} \label{App_md_methods}

The crystallization process of the supercooled Lennard-Jones liquid is considered, where the interaction between particles is determined by the pair potential:
\begin{equation}
U(r)=4\epsilon\left[\left(\frac{\sigma}{r}\right)^{12}-\left(\frac{\sigma}{r}\right)^{6}\right].
\end{equation}
Here $r$ is the distance between particles, $\sigma$ is the effective diameter of a particle, $\epsilon$ is the parameter that characterizes the depth of the potential well. Units of physical quantities are expressed in the terms of the potential parameters $\sigma$ and $\epsilon$: the temperature $T$ in the unit $\epsilon/k_{B}$, the pressure $p$ in the unit $\epsilon/\sigma^{3}$, the kinetic rate factors $g^{+}$ and $g^{-}$ in the unit $\tau^{-1}$, where $\tau=\sigma\sqrt{m/\epsilon}$ is the time unit, $m$ is the particle mass, $k_{B}$ is the Boltzmann constant. For argon with the parameters $m=6.63\times10^{-26}$~kg, $\sigma=0.341$~nm and $\epsilon/k_{B}=119.8$~K we have the value $\tau\simeq2.15$~ps. 

The considered system contains $N=13500$ particles located inside the simulation cubic cell with periodic boundary conditions in all three directions. We use the standard velocity-Verlet integrator and the time-step $\Delta t=0.01~\tau$ for integrate Newton's equations of motion and calculate the trajectories of particles~\cite{Verlet_1967}. The simulation was performed in the isothermal-isobaric ensemble. Pressure and temperature were controlled via the Nose-Hoover barostat and thermostat, respectively. The damping thermostat and barostat constants were taken to be $Q_{T}=100\Delta t$ and $Q_{p}=1000\Delta t$, respectively. These values of the quantities $Q_{T}$ and $Q_{p}$ are optimal for the system at the considered ($p$,$T$)-state. The initial system is a crystal with the face-centered cubic lattice at the temperature $T=0\,$K. Further, the system was heated to the temperature $T=2.5\,\epsilon/k_B $ at the constant pressure $p=2.0\,\epsilon/\sigma^3$ and, then, the system was brought to equilibrium. To prepare a supercooled sample, the equilibrated melt was rapidly cooled with the rate $0.04\,\epsilon/(k_{B}\tau)$ to the temperature $T=0.5\,\epsilon/k_B$. This temperature corresponds to the supercooling level $\Delta T/T_{m}=0.43$, where $\Delta T=T_{m}-T$ and the melting temperature is $T_{m}=0.88\,\epsilon/k_B$ on the isobar $p=2.0\,\epsilon/\sigma^3$. The numerical density $\rho$ of the system at this pressure is $\simeq0.92\,\sigma^{-3}$ [see phase diagram of the system in Ref.~\cite{LJ_PD}]. Detection of structural changes in the system starts immediately after receiving a supercooled liquid state. 

The centers of the crystalline phase are identified using cluster analysis based on estimation of the local orientational order parameters~\cite{Steinhardt_1983}. Particles involved in the formation of the crystalline structure are detected by ten Wolde-Frenkel condition~\cite{Wolde_Frenkel_1996}
\begin{equation}\label{WF_condition}
0.5<\left|\sum_{m=-6}^{6}\bar{q}_{6m}(i)\bar{q}_{6m}^{*}(j)\right|\leq1,
\end{equation}
where
\begin{equation}
\bar{q}_{6m}(i)=q_{6m}(i)\bigg/\sqrt{\sum_{m=-6}^{6}\left|q_{6m}(i)\right|^{2}}.
\end{equation}
Here, the $6$-fold bond order parameter is calculated through the expression
\begin{equation}
q_{6m}(i)=\frac{1}{n_{b}(i)}\sum_{j=1}^{n_{b}(i)}Y_{6m}(\theta_{ij},\phi_{ij}),
\end{equation}
where $Y_{6m}(\theta_{ij},\phi_{ij})$ are the spherical harmonics with the polar $\theta_{ij}$ and azimuthal $\phi_{ij}$ angles, $n_{b}(i)$ is the number
of neighbors for the $i$th particle. According to condition (\ref{WF_condition}), the $i$th particle is considered as involved in a crystalline phase if this particle has four or more crystal-like bonds with their own ``neighbors''. The most probable growth trajectory -- the time dependence of the nucleus size $n(t)$ -- of the largest crystalline nucleus is determined through the statistic treatment of the cluster analysis results. Here the $50$ independent trajectories $n(t)$ were used. The values of the crystal nucleation characteristics such as the critical size $n_c$ and the waiting time $\tau_{c}$ of the critically-sized nucleus are estimated by the method of inverted averaging of nucleus growth trajectories~\cite{Mokshin_Galimzyanov_2012, Mokshin_Galimzyanov_PCCP_2017}.

\end{document}